\documentclass{statsoc}

\usepackage[utf8]{inputenc}
\usepackage[T1]{fontenc}

\usepackage[a4paper]{geometry}
\usepackage{graphicx}
\usepackage[textwidth=8em,textsize=small]{todonotes}
\usepackage{amsmath}
\usepackage{amsfonts}
\usepackage{natbib}
\usepackage{url} 
\usepackage{bbm}

\usepackage{algorithmicx}
\usepackage{algpseudocode}
\usepackage{algorithm}

\title[Parameter Estimation for Dynamic Queueing Networks]{Likelihood-free parameter estimation for dynamic queueing networks: case study of passenger flow in an international airport terminal}
\author{Anthony Ebert}
\address{
	Queensland University of Technology, Brisbane, Australia
}
\email{ac.ebert@qut.edu.au}

\author{Ritabrata Dutta}
\address{
	University of Warwick, Coventry, UK
}

\author{Kerrie Mengersen}
\address{
	Queensland University of Technology, Brisbane, Australia
}

\author{Antonietta Mira}
\address{
	Università della Svizzera italiana, Lugano, Switzerland \newline
	and Università dell’Insubria, Como, Italy 
}

\author{Fabrizio Ruggeri}
\address{
	CNR-IMATI, Milano, Italy \newline
	and Queensland University of Technology, Brisbane, Australia
}

\author[A. Ebert, R. Dutta, K. Mengersen, A. Mira, F. Ruggeri and P. Wu]{Paul Wu}
\address{
	Queensland University of Technology, Brisbane, Australia
}

\renewcommand{\vec}[1]{\mathbf{#1}}

\newcommand*{\discrep}{\rho}

\newcommand{\muSyn}{1.47}
\newcommand{\sigmaSyn}{0.808}
\newcommand{\ratemgSyn}{1.51}
\newcommand{\ratesgSyn}{0.811}

\newcommand{\muReal}{1.18}
\newcommand{\sigmaReal}{0.694}
\newcommand{\ratemgReal}{1.27}
\newcommand{\ratesgReal}{1.66}

\newcommand{\ratemgTime}{47.2}
\newcommand{\ratesgTime}{36.1}

\newcommand{\muRealLow}{0.858}
\newcommand{\muRealHigh}{1.57}

\newcommand{\WalkLow}{27.2}
\newcommand{\WalkMed}{58.8}
\newcommand{\WalkHigh}{170}

\begin{document}

\begin{abstract}
Dynamic queueing networks (DQN) model queueing systems where demand varies strongly with time, such as airport terminals. With rapidly rising global air passenger traffic placing increasing pressure on airport terminals, efficient allocation of resources is more important than ever. Parameter inference and quantification of uncertainty are key challenges for developing decision support tools. The DQN likelihood function is, in general, intractable and current approaches to simulation make likelihood-free parameter inference methods, such as approximate Bayesian computation (ABC), infeasible since simulating from these models is computationally expensive. By leveraging a recent advance in computationally efficient queueing simulation, we develop the first parameter inference approach for DQNs. We demonstrate our approach with data of passenger flows in a real airport terminal, and we show that our model accurately recreates the behaviour of the system and is useful for decision support. Special care must be taken in developing the distance for ABC since any useful output must vary with time. We use maximum mean discrepancy, a metric on probability measures, as the distance function for ABC. Prediction intervals of performance measures for decision support tools are easily constructed using draws from posterior samples, which we demonstrate with a scenario of a delayed flight. 
\end{abstract}


\section{Introduction} \label{sec:intro}
Worldwide, air passenger numbers are expected to rise from 4.0 billion per year (2017) to 8.2 billion per year (2037) \citep{iata_iata_2018}. More passengers lead to longer queues and waiting times. Literature focused on modelling passenger flows within airport terminals is summarised in reviews by \citet{tosic_review_1992} and \citet{wu_review_2013}. These models aim to support decision makers in, cost-effectively, decreasing queues and waiting times. 

Significant seasonal and between-flight stochasticity exists in passenger demographics, and this induces variation in demand placed on airport terminals. These sources of variability together with ever-increasing security and immigration screening requirements, changing staff levels, baggage processing, and flow-on effects from other terminals increase the complexity of decision making. If decision makers are to use models to support decisions at real-life terminals, it is essential to provide not only accurate forecasts but also accurate assessments of uncertainty \citep{sacha2016role}.

None of these studies, however, considers the added contribution of parameter uncertainty or has a methodology for parameter estimation. In practice, parameters and their associated uncertainty are often difficult to estimate and are typically handled on a case-by-case basis. There are three main reasons for this. First, passenger flows in airport terminals represent a kind of dynamic queueing network (DQN), which are known to have intractable likelihoods \citep{insua_bayesian_2012}. Second, likelihood-free parameter estimation methods require large numbers ($> 10^5$) of simulations and, for such complex models, simulation times are long. Third, airport data are difficult to obtain, for academics as well as stakeholders, with numerous regulatory, commercial and technical challenges to be overcome. The data are never complete, so we must, therefore, work with what is available. 

\begin{figure}
	\centering
	\includegraphics[width=1\linewidth]{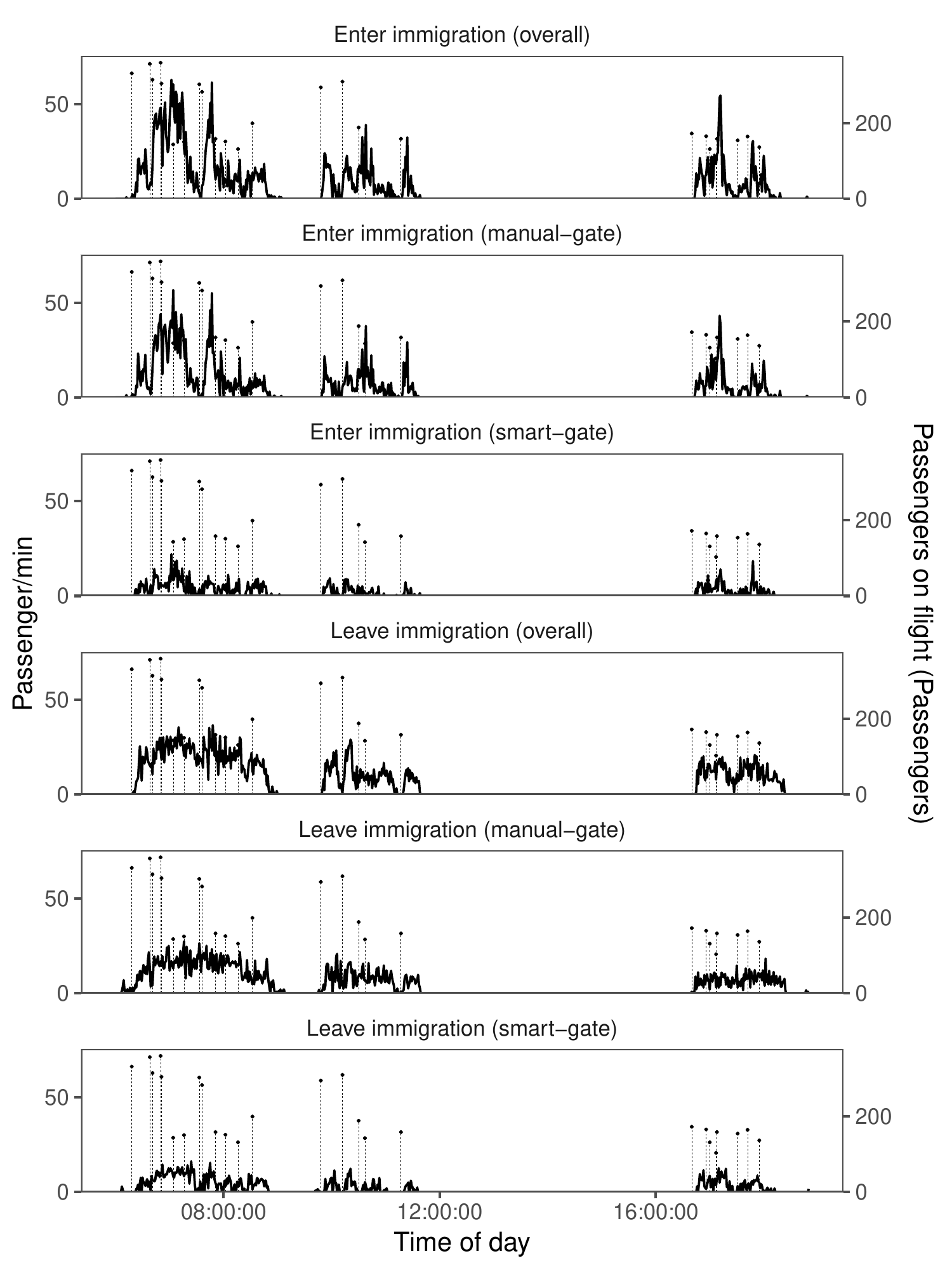}
	\caption{Solid lines: Observed passenger counts of flows within the arrivals terminal of an international airport terminal. Dashed vertical lines: Number of passengers on each flight arranged by flight arrival time. }
	\label{fig:airport_data}
\end{figure}

Motivated by passenger flows in airport terminals we develop a novel DQN inferential framework, enabled by a recent advance in queueing simulation speed \citep{ebert_computationally_2017}, called queue departure computation (QDC). Computational speed-ups, of more than two orders of magnitude, make simulation-based inference approaches, such as approximate Bayesian computation (ABC), feasible for a large DQN such as an airport terminal. The data we have available are of passenger counts per minute passing through different parts of an airport. The nature of these data (Figure \ref{fig:airport_data}) leads us to repurpose maximum mean discrepancy (MMD), a metric developed for probability measures, as a distance between observed data and model realisations.

The DQN model we construct has two main purposes: the first is to simulate model realisations, conditional on proposed parameter values, closely matching observed data so that inference may be made on these parameter values with ABC; the second is to support decision-makers in managing the airport terminal. The model output required for each purpose is different. In the first case, we are interested in generating passenger flow counts to compare with observed flow counts. In the second case, we are interested in generating output, with corresponding assessments of uncertainty, relevant to airport management (performance measures), such as queue lengths and waiting times. 

The paper is structured as follows. Section \ref{sec:queue} summarises the methodology and notation of queueing theory. In Section \ref{sec:PE_DQN} we introduce our modelling framework and explain how we intend to use ABC with MMD for DQN parameter inference. We provide some background for ABC in Section \ref{sec:ABC}. In Section \ref{sec:application} we construct a detailed simulation model of an airport passenger terminal, show how to proceed with parameter inference for real and synthetic data, and construct prediction intervals for performance measures of interest to inform decision making. In Section \ref{sec:discussion} we summarise our findings, discuss the advantages and the shortcomings of our approach and propose future work. 

\section{Queueing theory} \label{sec:queue}

A queueing system can be used to model any process where we can make the analogy to the queues we encounter in our everyday experience, namely, customers waiting in a line to be served by a server. We say analogy since the term customer could refer to: a web-query \citep{sutton_bayesian_2011}, a patient in a hospital \citep{takagi_queueing_2016}, a shipping container in a seaport \citep{kozan_comparison_1997}, an item in a manufacturing system \citep{dallery_manufacturing_1992}, or phone calls \citep{gans_telephone_2003}. Similarly, server could refer to a web-server, medical staff, machinery or a customer service representative at a call centre. Each customer $j = 1,2,\cdots$ arrives to the system at a certain arrival time $a_j$ and requires time $s_j$ with a server, called the service time. Vectors of arrival and service times, ordered by customer, are denoted as $\vec{a} = (a_1, a_2, \cdots)$ and $\vec{s} = (s_1, s_2, \dots)$ respectively. Typically a server can serve only one customer at a time; a server which is currently serving another customer is said to be unavailable while a server without a customer is available. If all servers are unavailable when a customer arrives then the customer must wait in the queue until a server is available. The waiting times are denoted as $\vec{w} = (w_1, w_2, \dots)$. Once a customer has been served they depart the system, so there is a corresponding vector of departure times $\vec{d} = (d_1, d_2, \dots)$. Clearly $d_j = a_j + w_j + s_j$; however this simple formula belies the complexity of the simulation problem since $w_j$ is a non-trivial function of $\vec{a}$ and $\vec{s}$ which depends on the class of queueing system considered. We use the term queueing system interchangeably with queue for brevity, although a queue (collection of waiting customers) is only one component of a queueing system. 

Queueing systems are classified according to a set of criteria introduced by \citet{kendall_stochastic_1953}. A queueing system is denoted by $f_\delta / f_{\vec{s}} / K / C / n / R$ where $f_\delta$ is the distribution of inter-arrival times $\delta_j = a_j - a_{j-1}$; $f_{\vec{s}}$ is the distribution of service times; $K$ is the number of servers; $C$ is number of customers that the system can hold (in the queue or currently in service); $n$ is the total number of customers and $R$ denotes the way that customers in the queue are allocated to servers, referred to as the queue discipline. The most common example of a queueing system has exponential ($M$) inter-arrival and service times, with one server, infinite system capacity, infinite number of customers and a first-come-first-serve (FCFS) queue discipline. In the notation of \citeauthor{kendall_stochastic_1953}, this is a $M / M / 1 / \infty / \infty / FCFS$ queue, almost always shortened to $M / M / 1$. The inter-arrival and service times in the $M / M / 1$ queueing system are drawn independently from exponential processes $\delta \sim \exp(\lambda_{\delta}), \vec{s} \sim \exp(\lambda_{\vec{s}})$, where $\lambda_{\delta}$ and $\lambda_{\vec{s}}$ are the rate parameters for each exponential distribution respectively. Another common distribution class for $f_{\vec{\delta}}$ or $f_{\vec{s}}$ is general independent (G) where inter-arrival or service times are iid samples from arbitrary distributions. 


Early work on queueing theory derive results mapping a queue's classification and parameters such as $\lambda_{\delta}$ and $\lambda_{\vec{s}}$ to steady-state distributions of performance measures such as: the number of customers in the system; the number of busy servers; customer waiting times; and the queue length. Bayesian posterior distributions are derived for: M/M/1 queues \citep{armero_bayesian_1994}; M/M/K queues \citep{wolff_problems_1965}; and M/G/1 queues \citep{wiper_mixtures_2012} as well as many other queueing systems \citep{insua_bayesian_2012}. 

A serious difficulty in Bayesian inference of queueing systems is obtaining the likelihood for a particular data collection scheme \citep{armero_dealing_1999}. The likelihood
function can be unavailable or very difficult to derive and therefore likelihood-free methods may be required \citep{insua_bayesian_2012}. The M/G/1 queueing model is widely studied within the literature of likelihood-free inference \citep{heggland_estimating_2004,blum_non-linear_2010,fearnhead_constructing_2012}.  However, to our knowledge, likelihood-free methods have never been used to study queues more complex than this single time-invariant system.

In a network of queueing systems \citep{jackson_networks_1957}, termed a queueing network (QN),
customers transition between queueing systems. After customers finish service at one
queueing system, they are assigned to their next queueing system. New customers
may enter from outside the system, and others leave the system entirely. Complex systems such as hospitals \citep{takagi_queueing_2016}, web-servers \citep{sutton_bayesian_2011} and biomolecular pathways \citep{ogle_proteolytic_2016} can all be modelled as QNs. 

A review of previous works of Bayesian inference for QNs can be found in \citet{armero_dealing_1999}. \citet{sutton_bayesian_2011} build a sophisticated Gibbs Sampler to derive posterior distributions $\pi(\theta|\mathbf{(a,d)})$ for a tandem QN of type $G/G/K/\infty/\infty$. Their technique applies to QNs where the arrival rate does not vary with time. Furthermore, their sampling algorithm is tailored to a particular data collection scheme where $\vec{a}$ and $\vec{d}$ are observed directly with censoring. If there is any measurement error in these observations, it is unclear how the algorithm will perform since the technique relies on proposing unobserved values of $\vec{s}$ such that the observations are consistent. If there is any model error or contaminated observations, there may not be a set $\vec{s}$ which is consistent with the dataset. 

QNs, with varying arrival rates, are termed dynamic queueing networks (DQN) and are commonly used to model queues in airport terminals \citep{wu_hybrid_2014}, call centres \citep{brown_statistical_2005}, and hospitals \citep{armony_patient_2015}. \citep{brown_statistical_2005}, one of the very few works considering inference on DQNs, uses a frequentist parameter inference scheme for a dataset collected from a call centre where arrival, wait and service times are observed in full, and $f_{\delta}$ is an inhomogeneous Poisson process. Once a QN reaches a certain level of complexity the relationship of the input (arrival and service times) to output (departure times) involves temporal dependency structures of unknown duration leading to intractable likelihoods \citep[Chapter~7]{insua_bayesian_2012}, especially where arrival or service rates change with time as in a dynamic queueing network (DQN). 

\section{Parameter Inference for Dynamic Queueing Networks} \label{sec:PE_DQN}

In this work, we provide a general approach to parameter inference for DQNs. We estimate unknown parameters $\theta$ of DQNs by embedding a queueing simulator within  approximate Bayesian computation, a likelihood-free inference scheme. Traditional simulation methods for DQNs, such as discrete event simulation \citep{nance_time_1981}, are computationally expensive. This makes simulation-based inference schemes like ABC infeasible for large systems. However, the computational efficiency of QDC, the adopted method for simulating queues, makes such an approach feasible. 

\begin{figure}
	\centering
	\includegraphics[width=0.8\linewidth]{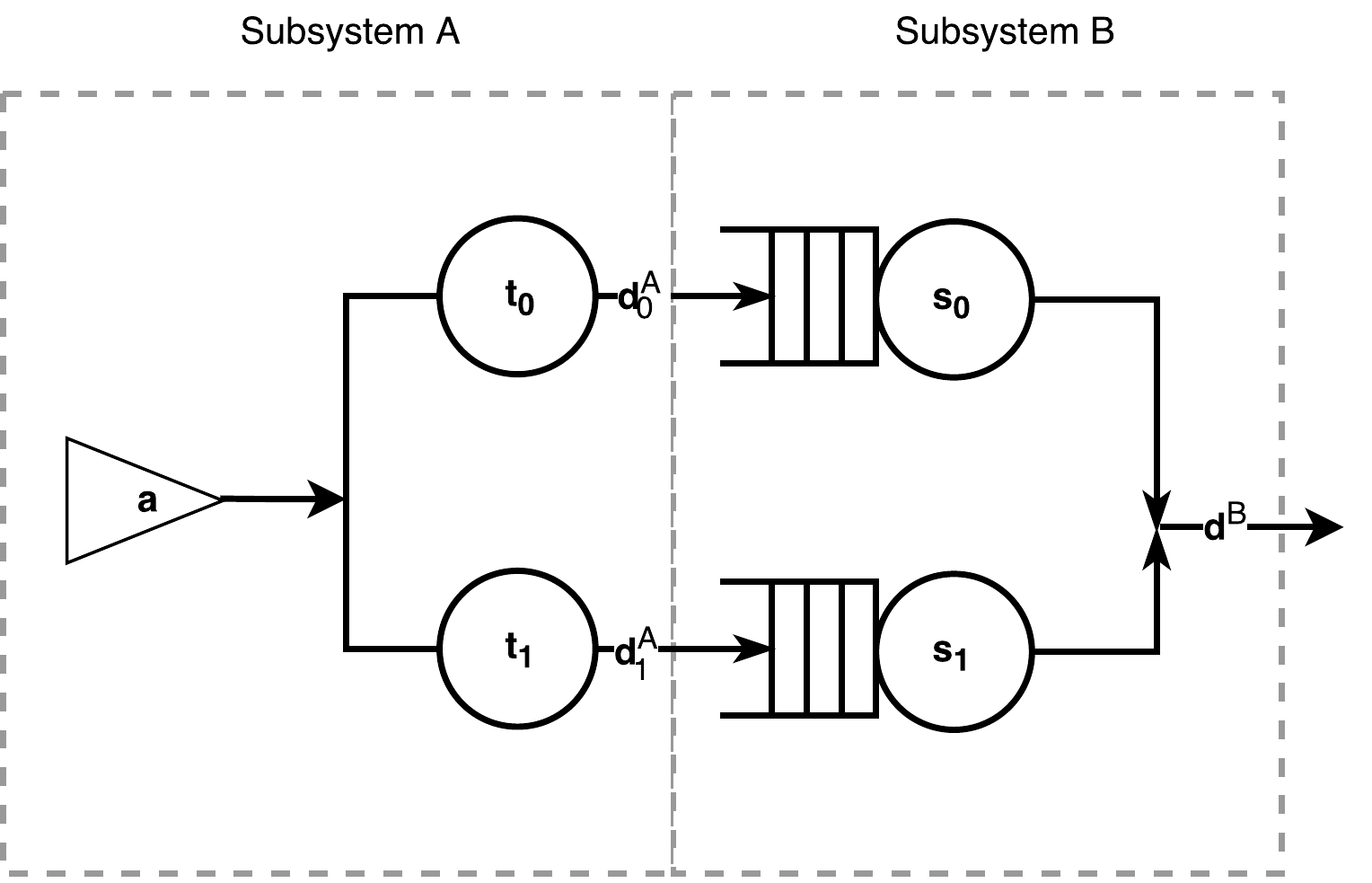}
	\caption{Illustrative example of a queueing network; here $\vec{t}_{r}$ denotes the subvector from $\vec{t}$ where the customer route is $r$.}
	\label{fig:play_example}
\end{figure}

As there are many sources of variation, the algorithm for simulating the system should be described in the language of a statistical model. We introduce our notation with an illustrative example (Figure \ref{fig:play_example}) of how a system with parallel queues could be simulated. In this example, customers must pass through traverse a number of stages which we refer to as subsystems. 

The system input $\vec{a}$, a vector of customer arrival times, is a draw from some known but arbitrary density function $f_{\vec{a}}$, which represents the schedule of customer arrivals. The elements of $\vec{a}$ need not be independent or identically distributed. They are a single high-dimensional draw from a distribution $f_{\vec{a}}$. We consider $f_{\vec{a}}$ to be a dynamic input since there is no requirement that the arrival rate of customers is constant over any interval of time.

Once customers arrive, each customer $i$ is routed to one of the two queueing systems, $r \in \{0,1\}$. The probability of customer $i$ being assigned to queueing system 0 is $p_i$: 
\begin{align*}
	r_i &\sim \text{Bern}(p_i).
\end{align*}
The transition time $t_i$ to the queueing system after the routing assignment is assumed to follow a gamma distribution, 
\begin{align*}
	t_{i} &\sim \text{Gamma}(\alpha, \beta).
\end{align*}
Hence the times that customers arrive at their respective queues are $d^A_i = a_i + t_i$ (the superscript denotes the subsystem). The service time for each customer is sampled from an exponential distribution with parameter $\lambda_0$ or $\lambda_1$ depending on their route,
\begin{align*}
	s_i | r_i &\sim \text{Exp}(\lambda_{r_i}).
\end{align*}
The number of servers in each system is denoted as $\vec{K}_0$ and $\vec{K}_1$ respectively. These values constitute the resource schedule $\vec{K}$, the resources available to the system, which are known non-negative integer-valued step functions over time. The departure times from subsystem $B$ are computed with two queueing system simulations since there are two routes (0 and 1). Let $\vec{d}_r^{A}$ denote the subvector from $\vec{d}^A$ corresponding to customers in route $r$. Similarly, $\vec{s}_r$ denotes the subvector of $\vec{s}$ corresponding to customers in route $r$. The customer ordering in $\vec{s}_r$ and $\vec{d}_r^A$ is preserved. This is important because, in general, arrival and service times are not independent. The subvectors of departure times from each queueing system $\vec{d}^B_{r}$ are computed deterministically conditional on $\vec{d}^A_{r}$, $\vec{s}_{r}$ and $\vec{K}_{r}$, using the queueing simulation algorithm QDC:
\begin{align*}
	\vec{d}^B_{r} = \text{QDC}(\vec{d}^A_{r}, \vec{s}_{r}, \vec{K}_{r}),
\end{align*}
where the first argument denotes input vector of arrival times, the second argument denotes the input vector of service times, and the final argument denotes the servers available. From this output, we can derive performance measures over time such as queue-lengths and waiting times (see Section \ref{sec:appdiscussion}).  

Given some observation $\vec{y}$ of the system, (which could consist of $\vec{d}^A$, $\vec{d}^B$, their subvectors by route, performance measures or some combination thereof), we would like to infer posterior distributions for $\theta$, which consists of $\alpha$, $\beta$, $\lambda_0$ and $\lambda_1$. Since $f_\vec{a}$ is dynamic, the output $\vec{y}$ may also be dynamic. 

In general, the likelihood function $f(\vec{y}|\theta ;f_\vec{a}, \vec{K} )$ cannot be evaluated, but the simulation scheme does allow us to generate model realisations $\vec{x}$, conditional on $\theta$ and known inputs $f_{\vec{a}}$ and $\vec{K}$. Next we explain how to sample from the approximate posterior distribution of $\theta$ using ABC. 

\section{Approximate Bayesian Computation} \label{sec:ABC}

We wish to sample from the posterior distribution of parameters $\theta$ given observations $\mathbf{y}$. Computational frameworks such as Markov chain Monte Carlo (MCMC) rely on the evaluation
of the likelihood function $f(\mathbf{y}|\theta)$. As explained before, the likelihood functions for DQNs cannot be evaluated, but we can efficiently simulate from the model using QDC. Here we propose ABC, a likelihood-free inferential framework \citep{lintusaari_fundamentals_2016} to infer the parameters of DQN.

In ABC, we jointly sample $(\theta,\vec{x})$ from a distribution proportional to $\pi(\theta)f_{\epsilon}(\mathbf{y}|\theta)$, where $f_{\epsilon}(\mathbf{y}|\theta)$ is an approximation to the likelihood function $f(\mathbf{y}|\theta)$:
$$f_{\epsilon}(\mathbf{y}|\theta) = \int f(\vec{x}|\theta) \mathbb{K}_{\epsilon} \{ \rho(\vec{x},\mathbf{y}) \} \text{d} \mathbf{x} ,$$
where $\rho(\vec{x},\mathbf{y})$ is a distance on the sample space and where $\mathbb{K}_{\epsilon}$ is a probability density function with a large concentration of mass near $\rho(\vec{x},\mathbf{y}) = 0$. To sample from this joint distribution, we first sample $\theta^*$ from the prior distribution $\pi(\theta)$ and then simulate a $\vec{x}$ from QDC using $\theta^*$ and finally accepting or rejecting $\theta^*$ depending on the probability $\mathbb{K}_{\epsilon} \{ \rho(\vec{x},\mathbf{y}) \}$.  Different ABC algorithms can be grouped following their choice of $\mathbb{K}_{\epsilon}$, being proportional to  $\mathbbm{1} \{\rho(\vec{x},\mathbf{y})<\epsilon\}$ in sequential Monte Carlo ABC \citep{sisson_sequential_2007} and population Monte Carlo ABC \citep{beaumont_adaptive_2009}; or proportional to  $\exp\{-\rho(\vec{x},\mathbf{y})/\epsilon \}$ in simulated annealing ABC (SABC) \citep{albert_simulated_2015}. Finally, in all of these ABC algorithms, we decrease $\epsilon \rightarrow 0$ at each iteration of the sequential algorithm, to improve the approximation of the likelihood function and hence to draw samples more representative of the true posterior distribution. An optimal choice of this decreasing sequence of $\epsilon$ gives us an accurate algorithm with a minimal loss in computational efficiency. Here we choose an efficient way of adapting $\epsilon$ proposed by \citet{albert_simulated_2015}, using ideas from non-equilibrium dynamics and simulated annealing \citep{kirkpatrick1983optimization}. The choice of a continuous density function rather than a discontinuous one for $\mathbb{K}_{\epsilon}$ and the efficient adaptation of $\epsilon$, help us to sample approximately from the posterior distribution while minimising the number of DQN simulations, empirically shown by \citet{albert_simulated_2015}, in comparison to the other ABC algorithms.   

A common practice in ABC literature is to define $\rho$ as the Euclidean distance between lower-dimensional summary statistics $S: \vec{x} \mapsto S(\vec{x})$, which, if sufficient, provide us with a consistent posterior approximation \citep{didelot_likelihood-free_2011}. As sufficient summary statistics are not known for most of the complex models, the choice of summary statistics remains a problem \citep{csillery_approximate_2010} and they have been previously chosen in a problem-specific manner \citep{blum_comparative_2013, fearnhead_constructing_2012, gutmann2018likelihood}. For DQNs, the observation $\mathbf{y}$ cannot be easily transformed into summary statistics $S(\mathbf{y})$ as there is a complex dependence structure \citep{sutton_bayesian_2011}, the system evolves with time, and the inputs to this system $f_{\vec{a}}$ and $\vec{K}$ can change. For instance, the operating hours and the schedule of a bus terminal may change in the future, but we would like to be able to repeat the inference procedure without reworking the summary statistics. Hence, here we consider constructing distances directly between data sets rather than between the extracted summary statistics. 

If we consider $\vec{x}$ and $\mathbf{y}$ as functions of time, we could use a distance between functions such as the L$^2$ norm as $\rho(\vec{x},\mathbf{y})$. In this work, we consider maximum mean discrepancy (MMD) as a distance between functional data, which is a metric on probability distributions with the same definition as the integral probability metric of \citet{muller_integral_1997} and can be shown to be equivalent to an L$^2$ norm between kernel density estimates. Distances between probability measures have recently been used in ABC, when $\vec{x}$ is a set of independent and identically distributed draws (e.g., MMD \citep{park2016k2}, Wasserstein distance \citep{bernton_inference_2017} and Kullback-Leibler divergence \citep{jiang2018approximate}). \citet{bernton_inference_2017} also extended the Wasserstein distance to time-series and demonstrated its use for a stationary queueing model. \citet{sriperumbudur_hilbert_2010} discuss, in detail, the relationships between these distances. 

In general, MMD can be used as a distance between probability density functions or positive valued functions which integrate to one. In our case we are interested in positive-valued functions which integrate to a fixed number since the number of passengers is known in advance (see Figure \ref{fig:airport_data}). We propose to use MMD to measure discrepancy between these functions, as they share this property with probability densities even though they are not probability densities.

\citet{gretton_kernel_2007} proposed a biased estimator of MMD, which is asymptotically consistent \citep{gretton_kernel_2012}. Computation of MMD estimator avoids the numerical instability associated with the integration of empirical distributions, involved in the computation of L$^2$ norms, as we compute the distance between functional datasets directly. 
The definition of $\hat{\discrep}_{\text{MMD}}(\vec{x},\vec{y})$, for two samples $\vec{x}$ and $\vec{y}$, is as follows \citep{gretton_kernel_2012}:
\begin{align}
	\hat{\discrep}_{\text{MMD}}(\mathbf{x},\mathbf{y}) &=  \frac{1}{m^2} \sum_{i=1}^m \sum_{j = 1}^m k(x_i, x_j) + 
	\frac{1}{n^2} \sum_{i=1}^n \sum_{j = 1}^n k(y_i, y_j) \label{eq:MMD} \\
	&\quad \quad - \frac{2}{mn} \sum_{i=1}^m \sum_{j = 1}^n  k(x_i, y_j), \nonumber 
\end{align}
where $m$ is the length of $\vec{x}$, $n$ is the length of $\vec{y}$ and $k$ is a kernel function. In this paper, we use the Gaussian kernel function $k(x,y) = \exp\{-(x-y)^2/(2\sigma^2_k)\} $, where $\sigma_k$ is a fixed tuning parameter. 

We now apply our general approach to likelihood-free inference for DQN models to passenger flow in an international airport terminal.

\section{Passenger Flow in an International Airport} \label{sec:application}

\begin{figure}
	\centering
	\includegraphics[width=1\linewidth]{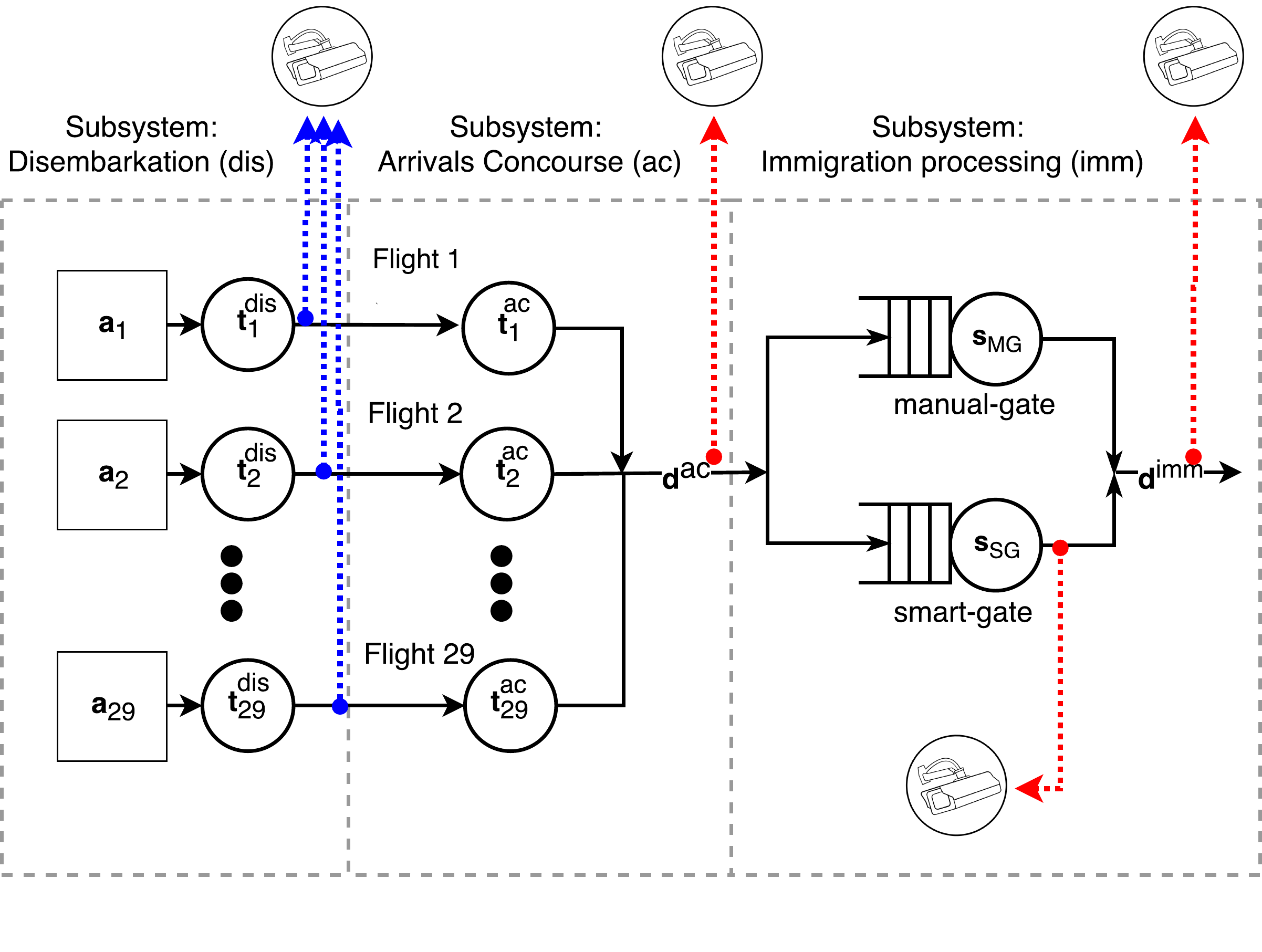}
	\caption{The facilitation process for arriving passengers at an international airport, visualised as a queueing network. The vertical blue lines represent CCTV-derived passenger count data used for the disembarkation parameters which were fitted before the ABC sampler. The red lines represent the CCTV-derived passenger counts used within the ABC sampler.}
	\label{fig:queuenet}
\end{figure}

The data comprise records, at each minute, of the number of persons entering and leaving a set of subsystems within the arrivals terminal (Figure \ref{fig:airport_data}). These measurements were derived from CCTV footage using the `virtual gate' algorithm of \citet{denman_large_2015}. The data have been slightly perturbed from the original data to anonymise them so that they may be made publicly available. There are many stages (subsystems) of passenger processing involved in the arrivals terminal (Figure \ref{fig:queuenet}). Firstly, passengers disembark from the arriving flight $i$ at the gate associated with that flight. 

The passenger then walks from their gate to the immigration sub-system to be processed. At this point, passengers take either the manual-gate (MG) or the smart-gate (SG) route through immigration subject to specific eligibility criteria such as nationality and age. After the passenger has been served at either the MG or the SG queueing system, they walk towards the baggage hall. This is the extent of our model. 

In addition to the CCTV-derived passenger counts, we have the associated flight schedule of arriving flights for that day. There are 29 flights and 5454 passengers in total. The flight schedule is a table with a set of information for each flight $i$ consisting of the arrival time $a_i$, the distance from the arrival gate to the immigration queue $m_i$, the number of passengers on the flight $j_i$, and the proportion of passengers who are local nationals $p^{\text{nat}}_{i}$ (as opposed to foreign nationals). Also supplied were the resource levels (resource schedule) assigned to each queueing system. This includes the number of machines at SG, $\vec{K}_{\text{SG}}$, and the number of staff members at MG, $\vec{K}_{\text{MG}}$, for each queueing system. The number of staff members assigned to MG changes with time according to the supplied staffing roster.

We compare the passenger counts for all subsystems with the information derived from the flight schedule $a_i$ and $j_i$ in Figure \ref{fig:airport_data}. The positions of the dashed vertical lines correspond to $a_i$, the heights to $j_i$ and the colours to $p^{\text{nat}}_{i}$. The black lines denote the passenger counts with time of day on the x-axis and number of passengers observed at that minute on the y-axis. We see that after the arrival of a flight there is a wave of passengers entering immigration. It is evident that it is not feasible to model this QN with a constant rate of passenger arrivals to immigration. The waves of passengers from flights overlap and we have only counts of passengers per minute so we cannot unambiguously identify the mix of flights that led to a particular inflow of passengers.

\subsection{Method} \label{sec:ex_sim} 

The aim is to obtain predictive distributions of performance measures given a flight and a resource schedule. This will allow planners to optimise resource allocation for a particular flight schedule. The flight schedule may represent a planned schedule in the future or an update to the current schedule with real-time information on flight delays. 

We model the system as a DQN as explained in the illustrative example in Section \ref{sec:PE_DQN}. The facilitation process (Figure \ref{fig:queuenet}) is divided into the following subsystems: disembarkation (dis), arrivals concourse (ac) and immigration processing (imm). The statistical model is constructed by generating a table of all passengers $j$ from every flight $i$ for the entire day of operation. We use statistical distributions to sample input variables such as disembarkation times $t^{\text{dis}}_{ij}$, walking times through the arrivals concourse $t^{\text{ac}}_{ij}$, route assignments $r_{ij}$ and service times at immigration $s_{ij}$. All variables with time units are denoted in minutes. We discuss now the statistical distributions used to model each stage of passenger processing. 

\begin{figure}
	\centering
	\includegraphics[width=1\linewidth]{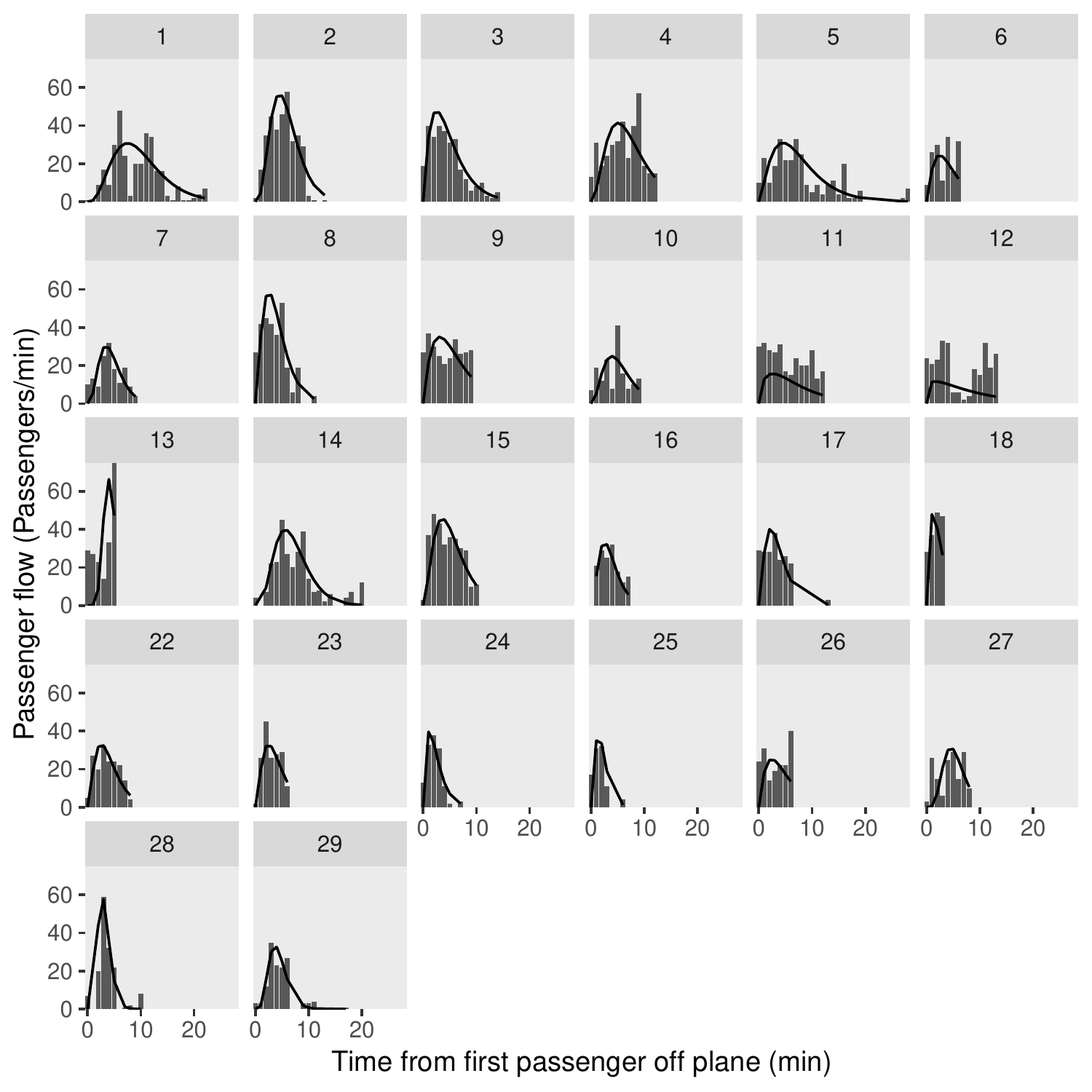}
	\caption{Passenger counts from CCTV cameras located at the arrival gates, organised by flight. The histograms denote the observed passenger counts after the flight has landed. The black lines represent the fitted gamma distribution for disembarkation. Disembarkation from the plane was much more time-consuming for most passengers than the walk from their aerobridge to immigration.  }
	\label{fig:disembark}
\end{figure}

The time taken by passenger $j$ from flight $i$ to disembark $t^{\text{dis}}_{ij}$ once the aircraft has landed is gamma distributed with flight-level parameters $\alpha^{\text{dis}}_i$ and $\beta^{\text{dis}}_i$ for shape and rate respectively, 
\begin{align*}
	t^{\text{dis}}_{ij} &\sim \text{Gamma} \left( \alpha^{\text{dis}}_i, \beta^{\text{dis}}_i \right). 
\end{align*}
We assume that along with the variability between flights there is also substantial variability in disembarkation profiles each day. We, therefore, consider these data insufficient for disembarkation modelling purposes and use robust frequentist parameter estimates based on the arrival gate passenger counts (Figure \ref{fig:disembark}). 

The time taken to walk from the arrival gate to immigration $t^{\text{ac}}_{ij}$ is gamma distributed with shape parameter $\alpha^{\text{ac}}$ and rate parameter $\frac{\beta^{\text{ac}}}{m_i}$, where $m_i$ is the distance in metres from the arrival gate of flight $i$ to the immigration queue and $\beta^{\text{ac}}$ can be interpreted as the rate parameter for a distance of 1 m, 
\begin{align*}
	t^{\text{ac}}_{ij} &\sim \text{Gamma} \left( \alpha^{\text{ac}}, \frac{\beta^{\text{ac}}}{m_i} \right).
\end{align*}
This is equivalent to simulating from $\text{Gamma}(\alpha^{\text{ac}}, \beta^{\text{ac}})$ and multiplying by $m_i$. We transform these parameters to mean $\mu^{ac} = \frac{\alpha^{\text{ac}}}{\beta^{\text{ac}}}$ and standard deviation $\sigma^{\text{ac}} = \frac{\sqrt{\alpha^{\text{ac}}}}{\beta^{\text{ac}}}$ parameterisation for interpretability, this allows us to compare results with \citet{al2007modeling}. The known parameters $m_i$ are not included in the reparameterisations since they vary by flight, but are still used within the simulation. 

Each customer is assigned a nationality $\text{nat}_{ij}$, which can be either local or foreign. This assignment is modelled as a Bernoulli variable with information from the known flight level parameters $p^{\text{nat}}_{i}$ (which is the probability that the passenger is local), 
\begin{align*}
	\text{nat}_{ij} &\sim \text{Bern}(p^{\text{nat}}_{i}). 
\end{align*}
The nationality of the passenger governs their propensity to take either route through immigration. The route assignments are Bernoulli variables and assign passengers to the SG route (as opposed to the MG route) with probability $p^{\text{imm}}_{\text{local}}$ or $p^{\text{imm}}_{\text{foreign}}$,
\begin{align*}
	r_{ij} | \text{nat}_{ij} &\sim \text{Bern}(p^{\text{imm}}_{\text{nat}_{ij}}).
\end{align*}
The service times $s_{ij}$ are exponentially distributed with rate parameter $\lambda_{\text{SG}}$ or $\lambda_{\text{MG}}$ depending on the route assignment, 
\begin{align*}
	s_{ij} | r_{ij} &\sim \text{Exp} \left( \lambda_{r_{ij}} \right).
\end{align*}
To predict the behaviour of the system for a future flight schedule, we concentrate on estimating unknown parameters which will not vary by day. In this case, the unknown parameters of interest are the walking parameters $\mu^{\text{ac}}, \sigma^{\text{ac}}$ and the service parameters $\lambda_{\text{SG}}$ and $\lambda_{\text{MG}}$. 

The final step is to use these simulated values to compute the times at which passengers transition between subsystems of the airport since this is the form of the observed data collected (Figure \ref{fig:airport_data}). For times at which passengers disembark and the leave arrivals concourse this is simple:
\begin{align*}
	d^{\text{dis}}_{ij} &= a_i + t^{\text{dis}}_{ij},  \\
	d^{\text{ac}}_{ij} &= d^{\text{dis}}_{ij} + t^{\text{imm}}_{ij},
\end{align*}
where $a_i$ is the time at which flight $i$ starts to allow passengers to deplane. Let $\vec{d}^{\text{ac}}$ be the vector of all $d_{ij}^{ac}$ and let $\vec{d}^{\text{ac}}_{r}$ be the subvector from $\vec{d}^{\text{ac}}$ of passengers in route $r$. The output of the immigration queueing system is computed with the queueing simulation algorithm QDC:
\begin{align*}
	\vec{d}^{\text{imm}}_{r}  &= 
	\text{QDC}(
	\vec{d}^{\text{ac}}_{r}, 
	\vec{s}_{r}, 
	K_{r}) &&\forall r \in \{\text{SG}, \text{MG} \},
\end{align*}
where $\vec{d}^{\text{imm}}_{r}$ is the subvector of departure times from the immigration system corresponding to customers in route $r$. 

We restructure the observed passenger counts corresponding to each subsystem $z$ and route $r$ into a vector $\widetilde{\vec{d}}^{z}_{r}$ to approximate a set of departure times from the QDC algorithm. Each minute of the day is repeated according to the number of passengers recorded at that minute of the day so that the length of the resulting vector is equal to the number of passengers observed within the CCTV data. The tilde $\sim$ is added since the vector is not ordered by passenger like $\vec{d}^{z}_{r}$ from the QDC algorithm. It is unordered, so departure times cannot be unambiguously assigned to passengers. Due to measurement error, censored intervals and the fact that individual passengers are not tracked through the system, the observed $\widetilde{\vec{d}}$ cannot be used to reconstruct the input variables directly. 

We use $\vec{x}^z_r$ and $\vec{y}^z_r$ in place of the observed data $\widetilde{\vec{d}}^{z}_{r}$ and simulated realisations $\vec{d}^{z}_{r}$. The number of passengers in the CCTV-derived passenger counts $\vec{y}^{\text{ac}}, \vec{y}^{\text{imm}}$, and $\vec{y}^{\text{imm}}_{\text{SG}}$ are 4866, 5249 and 1468 respectively. The distance $\rho$ used in the SABC algorithm is equal to 
\begin{align*}
\rho(\vec{x}, \vec{y}) &=  \hat{\discrep}_{\text{MMD}}(\vec{x}^{\text{ac}}, \vec{y}^{\text{ac}}) ~ + ~ \hat{\discrep}_{\text{MMD}}(\vec{x}^{\text{imm}}, \vec{y}^{\text{imm}}) ~ +  \\
&~ \hat{\discrep}_{\text{MMD}}(\vec{x}^{\text{imm}}_{\text{SG}}, \vec{y}^{\text{imm}}_{\text{SG}}).
\end{align*}
We could have used all the observed data available to us, but chose to use only three of the six observed histograms of passenger counts (Figure \ref{fig:airport_data}) to improve computational efficiency.

We have constructed the entire simulation model within the R programming language \citep{rstats} using the package queuecomputer \citep{ebert_computationally_2017}. The computational efficiency of the QDC algorithm means that simulating passenger flows in the arrivals terminal is very fast. We have recorded the time taken to simulate an entire day of passenger movements of 5,454 people in the terminal as $\sim$0.03 s. The ABC sampler is built in Python with the package ABCpy \citep{dutta_abcpy:_2017}, which is a modularised framework for building ABC samplers.

The value of $\sigma_k$ controls the standard deviation of the Gaussian kernel used to compute $\hat{\rho}_{\text{MMD}}$ (Equation \ref{eq:MMD}), we used a value of $\sigma_k = 20$ min, the reasoning was that two passengers from the same flight with the same characteristics should pass through subsystems within 20 minutes of each other. 

Vaguely informative priors for the parameters are imposed, namely $\alpha^{\text{ac}} \sim \text{U}(0, 10)$, $\beta^{\text{ac}} \sim \text{U}(0, 10)$, $\lambda_{\text{SG}} \sim \text{U}(0, 2.5)$ and $\lambda_{\text{MG}} \sim \text{U}(0, 2.5)$. The walking parameters priors correspond to a walking speed predictive distribution with an 80\% predictive interval of $(12.2 \text{ m\,min}^{-1}, 523 \text{ m\,min}^{-1})$, this range is vague for the available literature (see Section \ref{sec:appdiscussion}). The walking speed distribution is constructed to resemble that of \citet{al2007modeling}, who recorded pedestrian walking speeds from a large collection of video footage. The maximum of the service parameter prior support is chosen from previous work with industry partners: it corresponds to an average rate of 1 customer per 24 s which is a much higher rate than what is found in practice. 

\subsection{Results}

Before analysing the data, we evaluated the performance of the functional distance estimator by testing whether we can retrieve known and arbitrarily chosen parameters from synthetic data. We infer posterior distributions for input parameters: $\mu^{\text{ac}}, \sigma^{\text{ac}}, \lambda_{\text{SG}}, \lambda_{\text{MG}}$ given observations $\vec{y}$ comprising $\vec{y}^{\text{ac}}$, $\vec{y}^{\text{imm}}$, and $\vec{y}^{\text{imm}}_{\text{SG}}$. Firstly, to test the accuracy of the procedure, we generate synthetic data $\vec{y}_{\text{syn}}$ by setting input parameters to arbitrary values. In this case the parameters were set to $\mu^{\text{ac}} = 1.41$ min\,m$^{-1}$, $\sigma^{\text{ac}} = 0.8$ min\,m$^{-1}$, $\lambda_{\text{SG}} = 0.8$ min$^{-1}$, $\lambda_{\text{MG}} = 1.4$ min$^{-1}$. 

\begin{figure}
	\centering
	\includegraphics[width = 1\linewidth]{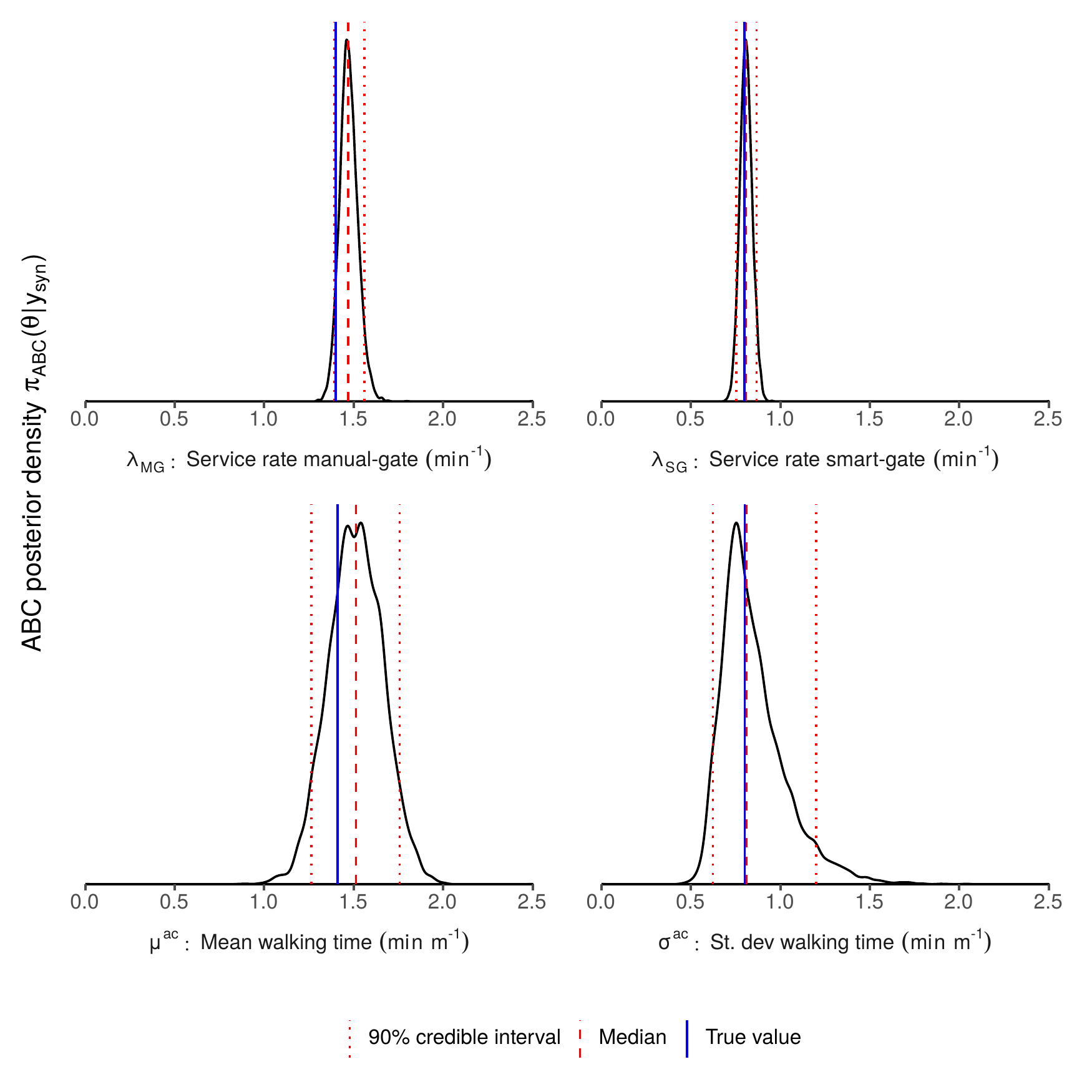}
	\caption{ABC posterior distributions based on synthetic data. The synthetic data was simulated with true values, shown as solid red vertical lines, $\mu^{\text{ac}} = 1.41$ min\,m$^{-1}$, $\sigma^{\text{ac}} = 0.8$ min\,m$^{-1}$, $\lambda_{\text{SG}} = 0.8$ min$^{-1}, \lambda_{\text{MG}} = 1.4$ min$^{-1}$, the posterior medians are $\muSyn{}$ min\,m$^{-1}$, $\sigmaSyn{}$ min\,m$^{-1}$, $\ratesgSyn{}$ min$^{-1}$ and $\ratemgSyn{}$ min$^{-1}$ respectively.}
	\label{fig:post_dist_syn}
\end{figure}

The ABC posterior distributions for the synthetic data (Figure \ref{fig:post_dist_syn}) show the relative performance of the ABC sampler in retrieving the true parameter. The vertical red line represents the true value for each parameter, and the vertical blue lines represent the posterior median and 90\% credible interval (CI). In all cases, the true value lies within the 90\% CI and close to the posterior median. The posterior variances of the service rate parameters are particularly small. 

\begin{figure}
	\centering
	\includegraphics[width = 1\linewidth]{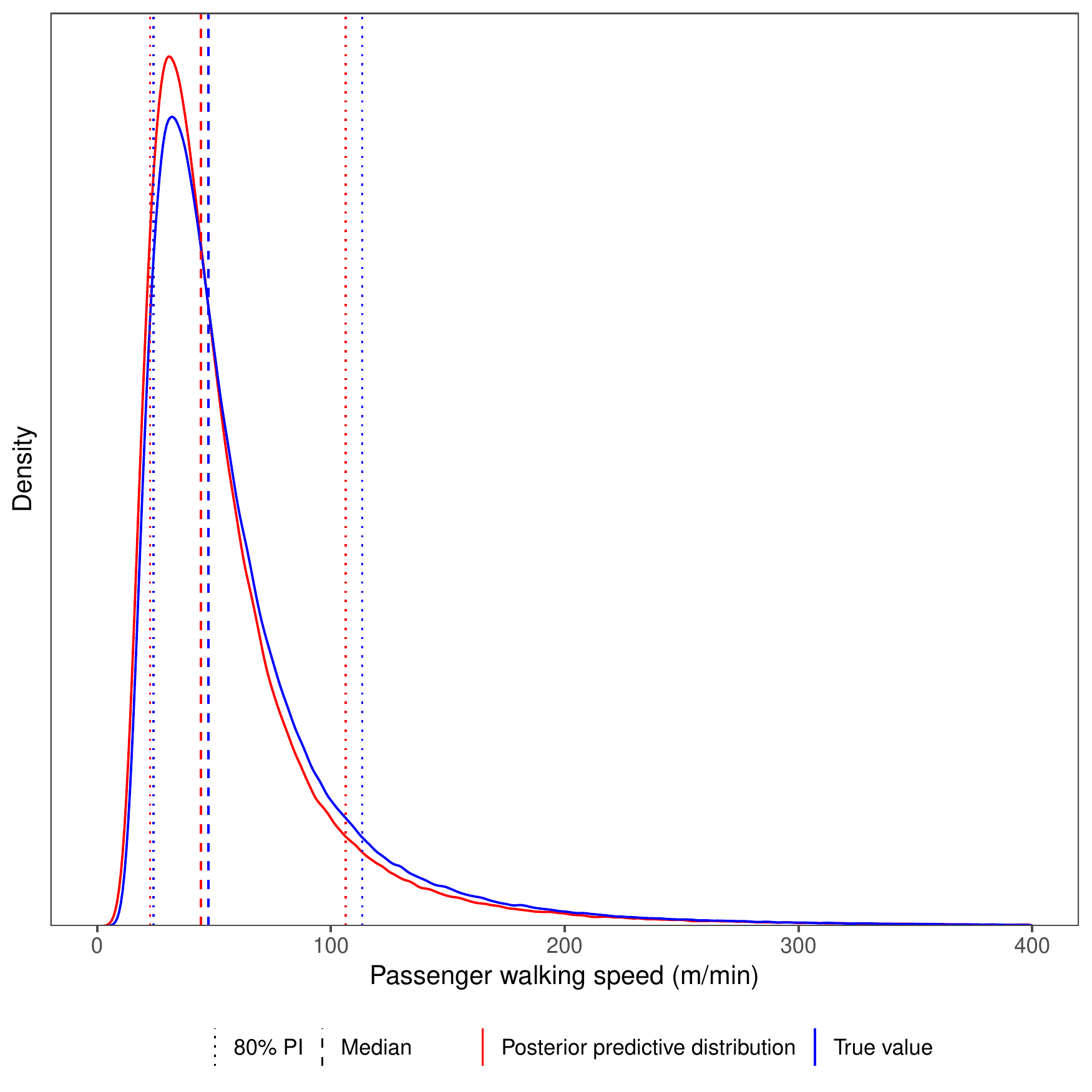}
	\caption{Posterior predictive distribution of walking speeds (synthetic data). To construct the posterior predictive distribution we draw samples of $\mu^{\text{ac}}$ and $\sigma^{\text{ac}}$ from $\pi_{\text{ABC}}(\theta | \vec{y}_{\text{syn}})$, then simulate $t_{ij}$ for one flight with $m_i = 1$, the walking speeds are equal to $\frac{1}{t_{ij}}$. We follow the same process, while fixing the parameter values to $\mu^{\text{ac}} = 1.41$ min\,m$^{-1}$, and $\sigma^{\text{ac}} = 0.8$ min\,m$^{-1}$ (the values used to simulate the synthetic data), to construct the ``true'' predictive distribution of walking speeds.} 
	\label{fig:walk_dist_syn}
\end{figure}

We obtained accurate and precise posterior distributions for all parameters. The posterior medians are \muSyn{} min\,m$^{-1}$, \sigmaSyn{} min\,m$^{-1}$, \ratesgSyn{} min$^{-1}$ and \ratemgSyn{} min$^{-1}$ for parameters $\mu^{\text{ac}}, \sigma^{\text{ac}}, \lambda_{\text{SG}}, \lambda_{\text{MG}}$ respectively. We computed the posterior predictive distribution of walking speeds and compared this to the distribution specified by the fixed and known values. The predictive distribution closely matched the true distribution of walking speeds (Figure \ref{fig:walk_dist_syn}). 

\begin{figure}
	\centering
	\includegraphics[width = 1\linewidth]{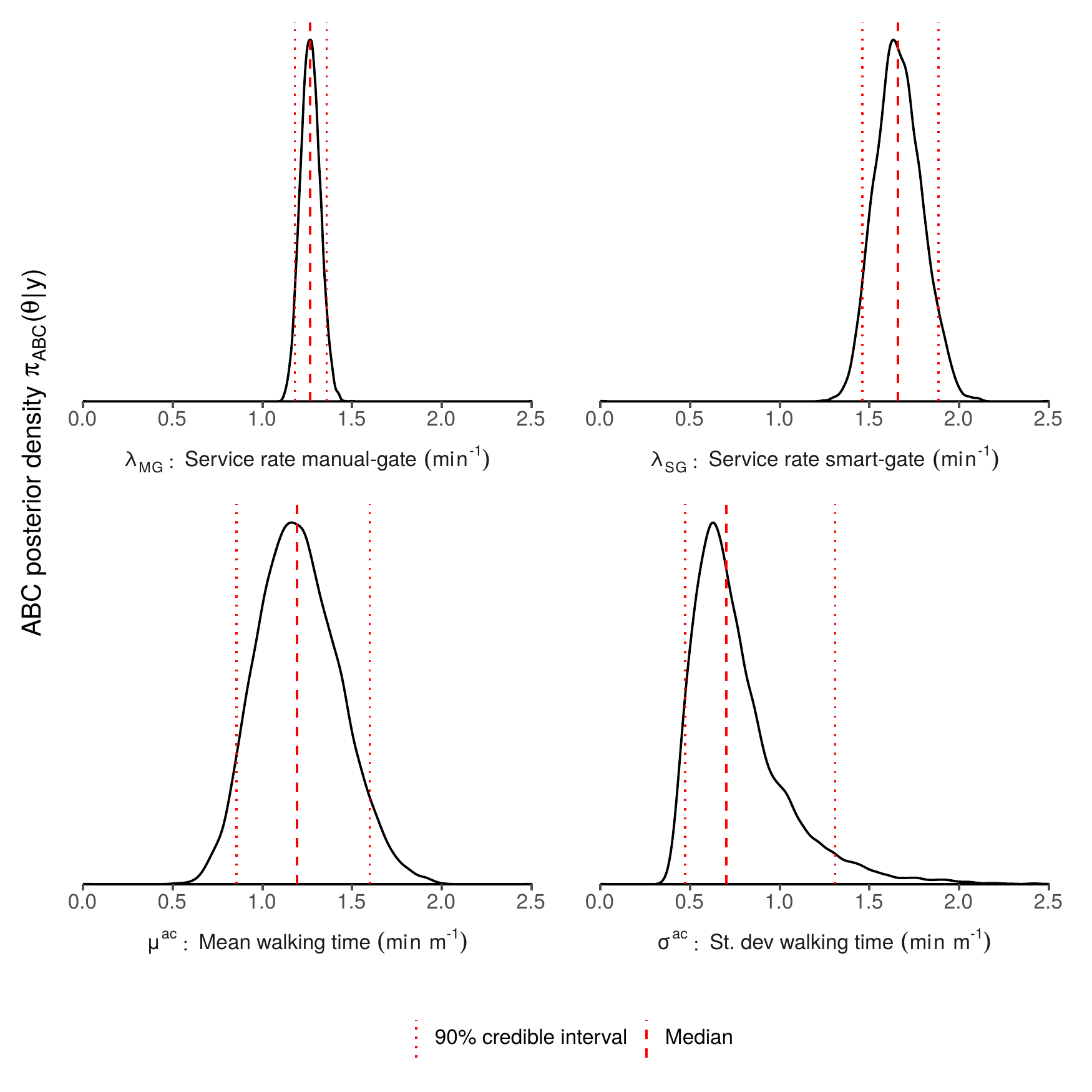}
	\caption{ABC posterior distributions based on real CCTV-derived passenger count data, as shown in Figure \ref{fig:airport_data}. The posterior medians are \muReal{} min\,m$^{-1}$, \sigmaReal{} min\,m$^{-1}$, \ratesgReal{} min$^{-1}$ and \ratemgReal{} min$^{-1}$ for parameters $\mu^{\text{ac}}, \sigma^{\text{ac}}, \lambda_{\text{SG}}, \lambda_{\text{MG}}$ respectively.}
	\label{fig:post_dist_real}
\end{figure}

We consider now the observed data. The ABC marginal posterior distributions are shown in Figure \ref{fig:post_dist_real}. The posterior medians are \muReal{} min\,m$^{-1}$, \sigmaReal{} min\,m$^{-1}$, \ratesgReal{} min$^{-1}$ and \ratemgReal{} min$^{-1}$ for parameters $\mu^{\text{ac}}, \sigma^{\text{ac}}, \lambda_{\text{SG}}, \lambda_{\text{MG}}$ respectively. The service rates correspond to median times of \ratesgTime{} s and \ratemgTime{} s for SG and MG customers respectively, which are reasonable. 

\begin{figure}
	\centering
	\includegraphics[width = 1\linewidth]{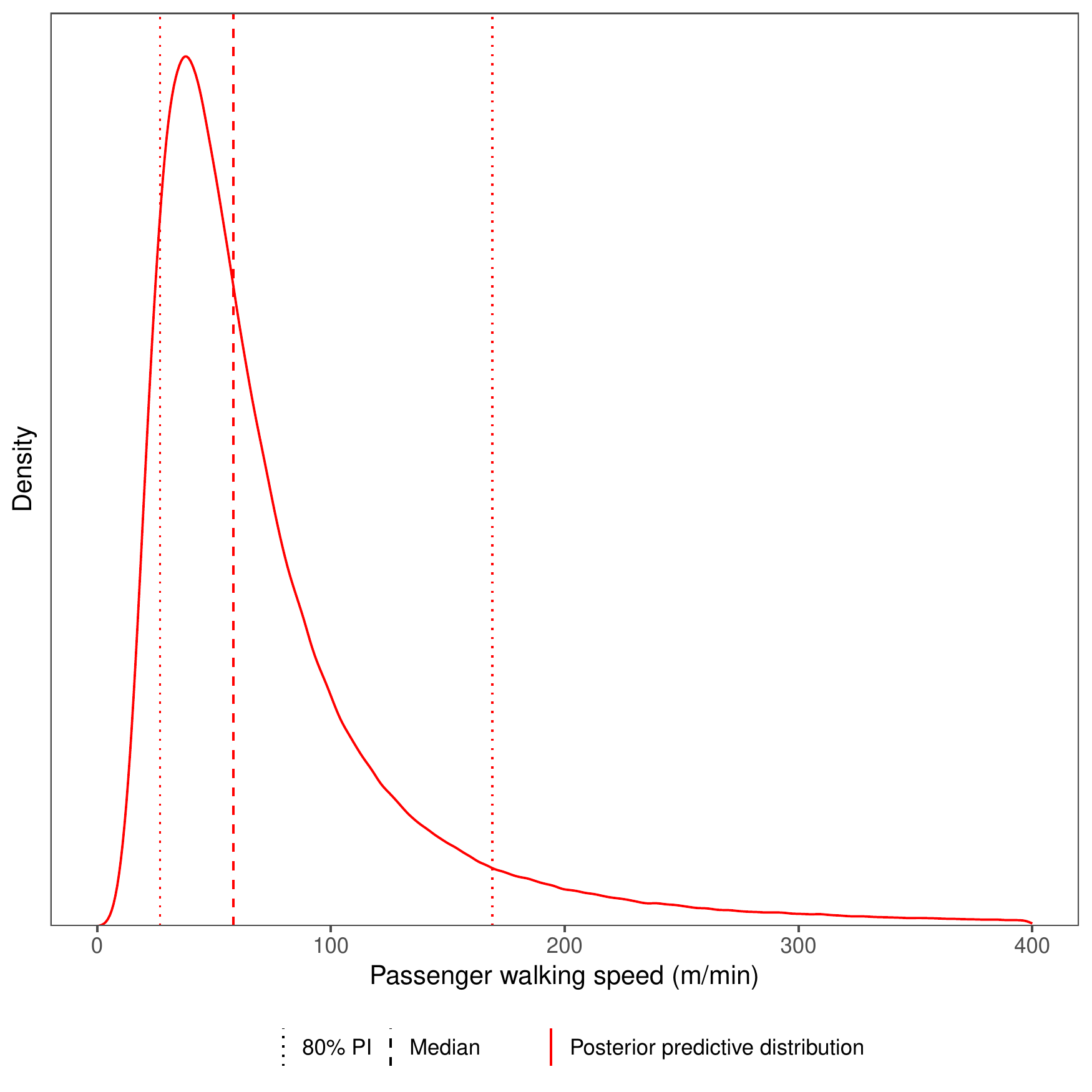}
	\caption{Posterior predictive distribution of walking speeds (real data). To construct the posterior predictive distribution we draw samples of $\mu^{\text{ac}}$ and $\sigma^{\text{ac}}$ from $\pi_{\text{ABC}}(\theta | \vec{y}_{\text{obs}})$, then simulate $t_{ij}$ for one flight with $m_i = 1$, the walking speeds are equal to $\frac{1}{t_{ij}}$. }
	\label{fig:walk_dist_real}
\end{figure}

We discuss now the walking parameters. Figure \ref{fig:walk_dist_real} shows the posterior predictive distribution for walking speed, obtained using the real data. The average reciprocal of speed (ROS) is  the parameter $\mu^{\text{ac}}$ with a posterior median of \muReal{} min\,m$^{-1}$ with 90\% CI $(\muRealLow{} \text{ min\,m$^{-1}$}, \muRealHigh{} \text{ min\,m$^{-1}$})$. The median posterior-predictive walking speed is \WalkMed{} m\,min$^{-1}$ with 80\% predictive interval (PI) $(\WalkLow{} $ m\,min$^{-1}$, $\WalkHigh{} $ m\,min$^{-1}$). 

\begin{figure}
	\centering
	\includegraphics[width = 0.9\linewidth]{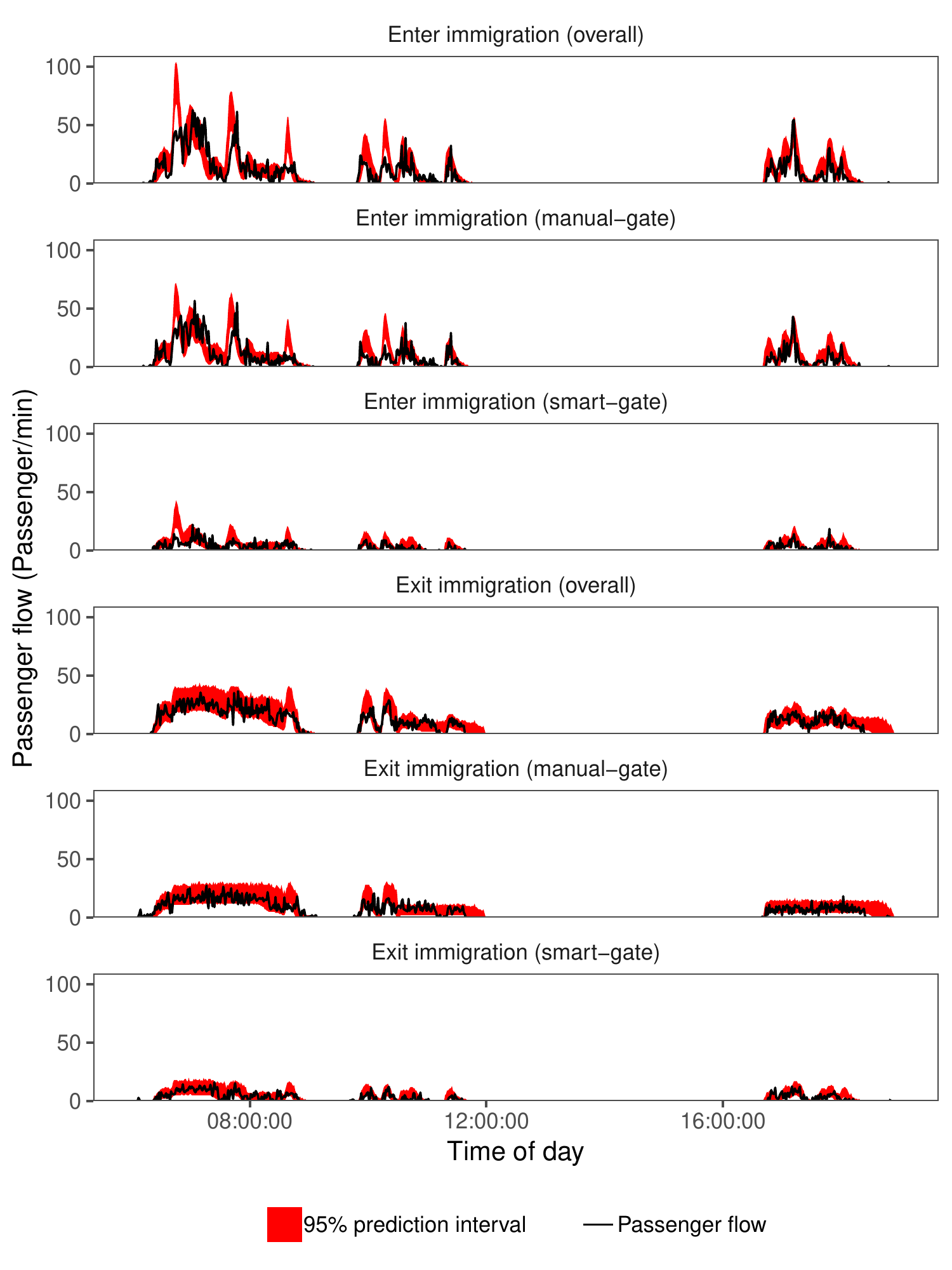}
	\caption{Comparison between passenger count data $\vec{y}$ for route combination for $\{MG, SG\}$ with corresponding 95\% posterior prediction intervals from model realisations $\vec{x}$.}
	\label{fig:model_realisations}
\end{figure}

Once we have approximations for the posterior distributions (Figure \ref{fig:post_dist_real}), we can draw $\theta \sim \pi(\theta|\vec{y})$ as an intermediary step to drawing $\vec{x}$ and deriving predictive intervals to compare with $\vec{y}$ (Figure \ref{fig:model_realisations}), so as to assess the validity of the model. 

\begin{figure}
	\centering
	\includegraphics[width = 1\linewidth]{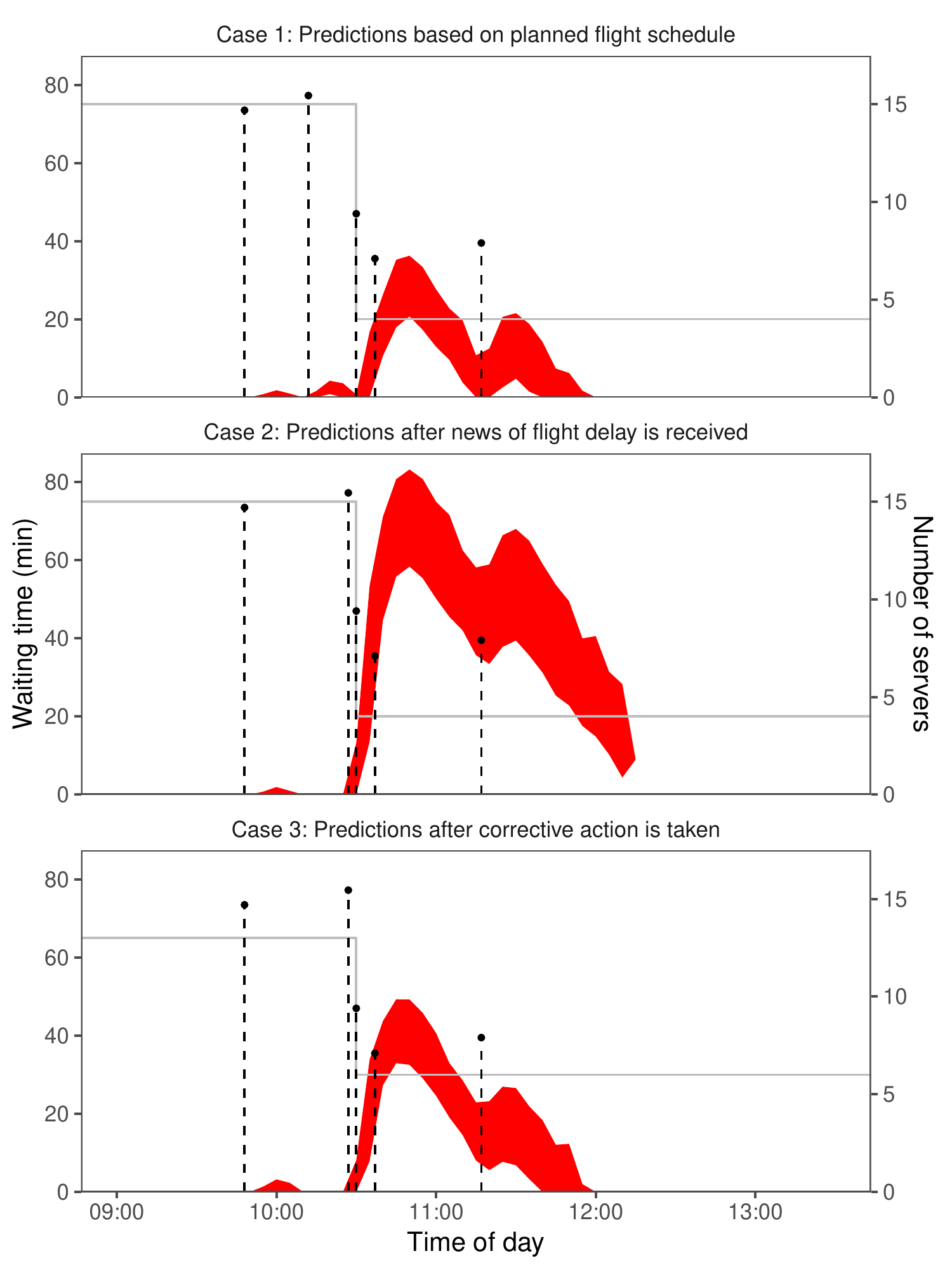}
	\caption{Waiting time predictions. Prediction intervals (95\%) for waiting times at the manual gates of immigration are shown as a red ribbon, and flight arrival times are indicated by the positions of the dashed vertical lines whose height is proportional to the number of passengers on that flight. The number of servers in each roster in indicated by the step function. Each plot shows different cases for the same scenario. In Case 1, we have the prediction based on the planned flight schedule. In Case 2, we have received news that the second flight is delayed by 15 minutes; this has a large effect on waiting times. In Case 3, we take corrective action by moving two servers from the earlier shift to the later shift.}
	\label{fig:wait_pf}
\end{figure}

\begin{figure}
	\centering
	\includegraphics[width = 1\linewidth]{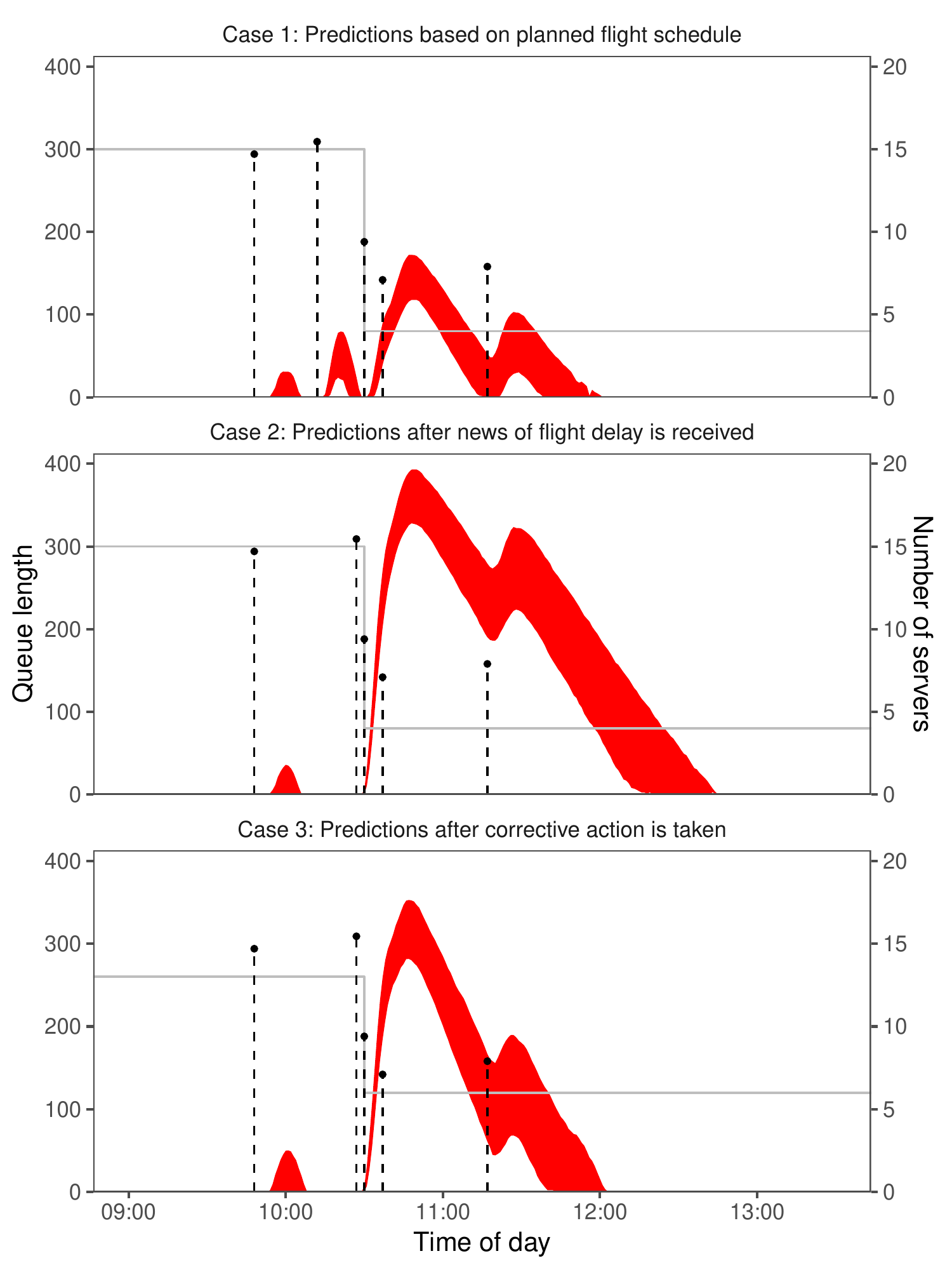}
	\caption{Queue length predictions. Prediction intervals (95\%) for queue lengths at the manual gates of immigration are shown as a red ribbon, and flight arrival times are indicated by the positions of the dashed vertical lines whose height is proportional to the number of passengers on that flight. The number of servers in each roster in indicated by the step function. The same scenarios are shown as in the previous plot, but we show the queue length rather than the waiting time.}
	\label{fig:queue_pf}
\end{figure}

The purpose of drawing ABC posterior samples for this model is to use those parameters to predict future performance measures as a decision support tool. This is possible even though these performance measures are not contained in the original dataset. In other words, queue lengths and waiting times were not recorded and are not recoverable from the original data. We demonstrate (Figure \ref{fig:wait_pf} and \ref{fig:queue_pf}) how posterior samples may be used in this manner with a short scenario supporting the decisions of a duty officer working for the immigration department. The flight schedule and staff roster form part of the observed dataset between 9 am and 1 pm, we use the real data to emphasise the point that situations, as we discuss, do arise. Each case of the scenario is simulated 500 times so that 95\% prediction intervals may be constructed for the performance measures: queue lengths and waiting times. Passengers are binned (5 minutes) by arrival time to immigration $(d^{\text{ac}}_{ij})$ to calculate the average waiting time for the simulation and case-study. The queue length of the immigration system is binned (1 minute) to calculate the maximum queue length in the period. At the beginning of the day (Case 1), predictions follow from planned flight schedule; however, later we receive information that the second flight is delayed by 15 minutes. This seemingly minor change has a large effect on waiting times and queue lengths (Case 2). The waiting time at 10:50 am (the peak in all cases) increases from a median value of 28 minutes to 71 minutes. Similarly, the queue length increases from a median value of 146 to 363 passengers. The decision we make, faced with these numbers, is to move two staff from the earlier shift to the later shift (Case 3). This corrective action also has a large effect on waiting times and queue lengths, in the reverse direction, the median waiting time is reduced to 41 minutes, and the median queue length is reduced to 319 passengers. 

\subsection{Discussion} \label{sec:appdiscussion}



Passenger flows within airport terminals are part of a complex and dynamic system. There are retail outlets, bathroom facilities, family groups and congested passenger flows. This is particularly relevant for the distribution of walk times since passengers do not walk independently of each other. Nevertheless, we find that our posterior predictive distribution for walking speeds (Figure \ref{fig:post_dist_real}) places much of its mass within a reasonable range of walking speeds  with the posterior median located at \WalkMed{} m\,min$^{-1}$. \citet{young_evaluation_1999} performed a study of pedestrian walking speeds in airport terminals and found the average walking speed to be $80.5$ m\,min$^{-1}$. However, our 80\% CI is equal to $(\WalkLow{}$ m\,min$^{-1}$, $\WalkHigh{}$ m\,min$^{-1}$) which is very different to \citet{young_evaluation_1999} who report a standard deviation of $15.9$ m\,min$^{-1}$. The tails of our distribution are much heavier than \citet{young_evaluation_1999} and \citet{al2007modeling} which is perhaps a consequence of inferring walking times indirectly through a congested airport rather than recording independent observations of walking speeds from footage. 

Regarding the comparison between observed passenger counts and model realisations from posterior samples (Figure \ref{fig:model_realisations}), with only four unknown parameters in the model and with only passenger counts from three of the six subsystems used in the distance $\rho$ we see a close match for all subsystems. Many but not all the peaks corresponding to waves on passengers in the real-data overlap with the prediction interval. The match can appear close with peaks having the same size and widths, but a small translation in peak positions can have a large effect on a  functional distance estimator. These ``translation errors'' may, unfortunately, be an inevitable property of these data; we have noticed substantial between-flight rather than within-flight variation of walking speeds. The crest of a wave of passengers walking to immigration can vary from the predictive distribution by a few minutes in opposing directions depending on the flight \textemdash even after we correct for the walking distance from arrival gate to immigration. Perhaps whether a flight of passengers walks quickly or slowly is related to the speed of the first passengers to leave the plane. 

We have shown that a flight schedule, staffing roster and posterior distribution can be used together to produce a predictive interval for performance measures of interest, in this case waiting times (Figure \ref{fig:wait_pf}) and queue lengths (Figure \ref{fig:queue_pf}). Drawing samples from the posterior, we can modify the flight schedule and/or the staffing roster so as to optimise the system according to the performance measure. Computing prediction intervals for waiting times and queue lengths from 500 simulations for three cases of a single scenario, each involving 1,091 passengers in an airport took 35 s in total on a standard desktop. 

Computational cost was a key consideration during implementation\footnote{Code available at \url{https://github.com/AnthonyEbert/AirportPassengerFlow}}. A naive implementation of MMD leads to expensive computations as the cost of computing scales with $O\{m n\}$. The observed data $\vec{x}$ is binned into intervals of one minute, and the number of these intervals is much lower than the number of passengers, $m$ (similarly for $\vec{y}$). It makes sense, therefore, to compute the equivalent value with a weighted MMD over the number of intervals\footnote{Code available at \url{https://github.com/AnthonyEbert/EasyMMD}}, equal to the number of minutes. This leads to sampling time for $\vec{x} \sim f(\cdot | \theta)$ being roughly equivalent to $\hat{\discrep}_{\text{MMD}}(\vec{y},\vec{x})$. 

The number of parallel queueing systems, equivalently the number of unique values of $r$, in this paper is two (MG and SG). If this number is increased to some number $n_r$ (keeping the size of $\mathbf{x}$, $m$, fixed), the effect on sampling time is minimal. There would be more queueing systems, but fewer customers in each and computation time for queueing systems scales linearly with the number of customers. The greatest problem when increasing $n_r$ is partitioning $\vec{a}$ and $\vec{s}$ into distinct routes before QDC computation. This problem is equivalent to sorting a vector of size $m$ containing $n_r$ distinct values, which is known to scale as $O\{m \log_2(n_r)\}$ \citep{katajainen1994sorting}, so computing time scales sublinearly with the number of parallel queueing systems. 

\section{Conclusion} \label{sec:discussion}


We have demonstrated a novel DQN parameter inference framework. The framework requires only that it is possible to simulate realisations $\vec{x}$ which resemble $\vec{y}$ and that the resource schedule is known. Innovations such as QDC, for faster simulation times and MMD, for straightforward and robust distance computation have made the ABC sampler's task easier but neither is fundamental to our approach. Instead, the contribution is to conceive of the observed dataset as functional data and furthermore to use MMD as the notion of distance between functional datasets. Moreover, to our knowledge, this is the first work to address parameter estimation of a DQN in a Bayesian manner. 

In contrast to the approach of \cite{sutton_bayesian_2011}, we have used an ABC sampler which allows us to easily and robustly adapt the algorithm to any observational scheme. The cost is that, like all ABC sampling algorithms, we will have endured information loss and our approximation to the posterior will be biased towards the prior. The sampler of \cite{sutton_bayesian_2011} could be extended to DQNs, in which case the complexity of the model as well as the size and complexity of the observed dataset will have to be taken into account to decide the best approach for the situation. 

We have shown with a real-world example of an airport that the technique can be used to infer parameter distributions. The method is straightforward to apply and simple to adapt with changes to the DQN model. In our case, we have limited the model to the first part of the arrivals terminal of an international airport, but conceptually  it could be extended further in a straightforward manner. We have seen that MMD performed well for a synthetic dataset generated with known parameter values. Furthermore, in the case of the real dataset, use of MMD in the ABC sampler led to realistic prediction intervals in the face of model error and incomplete information. Future work could involve adaptation of the distance measure, namely how it responds to model misspecification and in what manner the approximation to the posterior is affected. Alternatively one could encode the data two-dimensionally as a histogram, i.e. (time, passenger count), treat this as a time-series and use the metric developed by \citet{bernton_inference_2017}. A way of assessing the performance of a distance in the ABC sampler is required so that judgements can be made regarding the correct distance to use. For instance, \citet{bernton_inference_2017} compared estimators for MMD and the Wasserstein distance for a statistical problem with a tractable likelihood. The comparison was made by computing the Wasserstein distance between the true posterior and the ABC posterior. 

A question regarding tandem DQNs is whether it is best to estimate all parameters at once, as we have, or to use the tandem structure of the DQN to infer parameters one by one. For instance, we could have used the count-stream $\vec{d}^{\text{ac}}$ to infer distributions for the walking parameters and then subsequently to infer distributions of service-rate parameters with $\vec{d}^{\text{imm}}$. This may have advantages in terms of scalability for large numbers of parameters or large numbers of subsystems; however, it is unclear whether this would still result in a valid joint posterior distribution. 

\section*{Acknowledgements}

Thank you to Hamish Macintosh, HPC, Queensland University of Technology and Dr Marcel Schoengens, CSCS, ETH Zurich for help with high performance computing. The authors wish to acknowledge the support of the QUT High Performance Computing and Research Group (HPC) and the Swiss National Supercomputing Centre (CSCS). Thank you to Dr Bulukani Mlalazi for contributing code to compute predictive distributions of performance measures. 

This work was supported by the ARC Centre of Excellence for Mathematical and Statistical Frontiers (ACEMS). This work was funded through the ARC Linkage Grant “Improving the Productivity and Efficiency of Australian Airports" (LP140100282). Ritabrata Dutta was supported by the Swiss National Science Foundation Grant No. 105218163196 (Statistical Inference on Large-Scale Mechanistic Network Models). 

\bibliographystyle{rss}
\bibliography{main}

\appendix

\end{document}